\documentclass[aps,prb,twocolumn,showpacs,superscriptaddress]{revtex4}
\usepackage[dvips]{graphicx}
\usepackage{dcolumn}
\usepackage{bm}
\usepackage{xspace}
\usepackage{multirow}
\usepackage{natbib}
\usepackage{color}
\usepackage{ulem}

\def\Ed#1{\textcolor{red}{\sout{}}}

\begin{document}
\title{Spin Configuration and Scattering Rates on the Heavily Electron-doped Surface of Topological Insulator Bi$_2$Se$_3$}

\author{Z.-H. Pan}
\affiliation{Condensed Matter Physics and Materials Science Department, Brookhaven National Lab, Upton, NY 11973}
\author{E. Vescovo}
\affiliation{National Synchrotron Light Source, Brookhaven National Lab, Upton, NY 11973}
\author{A. V. Fedorov}
\affiliation{Advanced Light Source, Lawrence Berkeley National Laboratory, Berkeley, CA 94720}
\author{B. Sinkovic}
\affiliation{Department of Physics, University of Connecticut, Storrs, Connecticut 06269, USA}
\author{D. Gardner}
\author{S. Chu}
\author{Y.S. Lee}
\affiliation{Center for Materials Science and Engineering, Massachusetts Institute of Technology, Cambridge, MA 02139}
\author{G. D. Gu}
\author{T. Valla}
\affiliation{Condensed Matter Physics and Materials Science Department, Brookhaven National Lab, Upton, NY 11973}
\date{\today}

\begin{abstract}

Heavily electron-doped surfaces of Bi$_2$Se$_3$ have been studied by spin and angle resolved photoemission spectroscopy. Upon doping, electrons occupy a series of {\bf k}-split pairs of states  above the topological surface state. The {\bf k}-splitting originates from the large spin-orbit coupling and results in a Rashba-type behavior, unequivocally demonstrated here via the spin analysis. The spin helicities of the lowest laying Rashba doublet and the adjacent topological surface state alternate in a left-right-left sequence. This spin configuration sets constraints to inter-band scattering channels opened by electron doping.  A detailed analysis of the scattering rates suggests that intra-band scattering dominates with the largest effect coming from warping of the Fermi surface.

\end{abstract}
\vspace{1.0cm}

\pacs {74.25.Kc, 71.18.+y, 74.10.+v}

\maketitle

\pagebreak

Three-dimensional topological insulators (TIs) display  Dirac-like surface states with a distinctive chiral spin-structure in which the electronic spin of the topological surface state (TSS) is locked perpendicular to $\bf k$ in the surface plane.\cite{Fu2007a,Noh2008a,Hsieh2008,Zhang2009,Hsieh2009a,Xia2009,Chen2009,Pan2011c}
A consequence of this chiral spin-structure is that elastic backscattering, which would require a spin-flip process, is forbidden when time-reversal-invariant perturbations are present.\cite{Fu2007a}  TSSs are therefore promising candidates for spintronics and quantum 
computing applications. \cite{Biswas2010,Fu2009,Guo2010,Liu2009,Zhou2009,Chen2010,Wray2010a,Yazyev2010} 
Indeed, recent scanning tunneling microscopy (STM) experiments \cite{Roushan2009,ZhangSTM2009,Alpichshev2010,Seo2010,Hanaguri2010} have convincingly shown that, in topological systems, the backscattering is strongly suppressed, or completely absent, despite the presence of considerable disorder at the atomic scale. 

 A high degree of spin polarization of the TSS in the model system,  Bi$_2$Se$_3$ has been recently confirmed by spin and angle resolved photoemission spectroscopy (SARPES).\cite{Pan2011c,Jozwiak2011} 
Furthermore, ARPES studies have shown that the adsorption of various non-magnetic atomic or molecular species on the surface of this compound results in the electronic doping of the TSS and  partial filling of additional states.\cite{Benia2011,Bianchi2011,Valla2012a} These states resemble Rashba states in that they form pairs of parabolic-like bands, displaced in momentum and intersecting at new Dirac points located at the zone center, similar to the states observed on Au(111). \cite{LaShell1996,Muntwiler2004} Subsequent ARPES studies have indicated that such states also form when magnetic impurities are adsorbed on the Bi$_2$Se$_3$ surface.\cite{Valla2012a} These studies have also shown that the TSS is surprisingly insensitive to the presence of both non-magnetic and magnetic impurities in the low doping regime and that the scattering rates are affected only at higher doping, when the initially small and circular Fermi surface (FS) has sufficiently enlarged and acquired a significant hexagonal warping.\cite{Valla2012a}

We note that, if the doublet states are indeed Rashba states, the doped surface would be an interesting realization of a system where several inter-band spin-dependent channels would be available for scattering.\cite{Fedorov2002} Particularly important would then be to determine the spin configuration of the doublet states relative to the one of the TSS.
 
In this article, we present the SARPES studies of the surface electronic structure of Bi$_2$Se$_3$ doped with Rb, Cu and Fe. Our results demonstrate that the paired states are indeed Rashba states. The lowest Rashba doublet forms two FSs: the inner FS has a left-hand (L) spin helicity, parallel to the TSS\rq{}, while the outer one has a righ-hand (R) spin helicity, antiparallel to the TSS\rq{}. Furthermore, we found that at high doping levels, the adjacent TSS and outer Rashba state are considerably broadened, while the inner Rashba state displays a line-width comparable to that of the TSS on pristine surface. These observations suggest that intra-band scattering channels dominate and that the broadening is likely due to the modification in shape and size of each individual Fermi surface. 

ARPES studies were performed at the beam line 12.0.1 of the Advanced Light Source (ALS) and  U13UB of the National Synchrotron Light Source (NSLS), using a Scienta-SES100 and a Scienta-2002 electron analyzers, respectively. The photon energies used in these experiments range from 18.5 eV to 65 eV. The energy  and angle resolution was $\sim15$ meV and better than $ 0.2^{\circ}$, respectively.
SARPES experiments were performed at the beamline U5UA at NSLS using an Omicron EA125 electron analyzer, coupled to a mini-Mott spin polarimeter for the measurements of the {\it in-plane} component of the spin polarization. The spin-resolved data were recorded at 18.5 eV photon energy, with the sample kept at $\sim30$ K. The combined energy and angle resolution was approximately 35 meV and $0.5^{\circ}$, respectively. 
All samples were cut from the same bulk piece and cleaved in ultrahigh vacuum conditions (base pressure better than $5\times 10^{-9}$ Pa in both the ARPES and SARPES chambers).

\begin{figure}[t]
\begin{center}
\includegraphics[width=8cm]{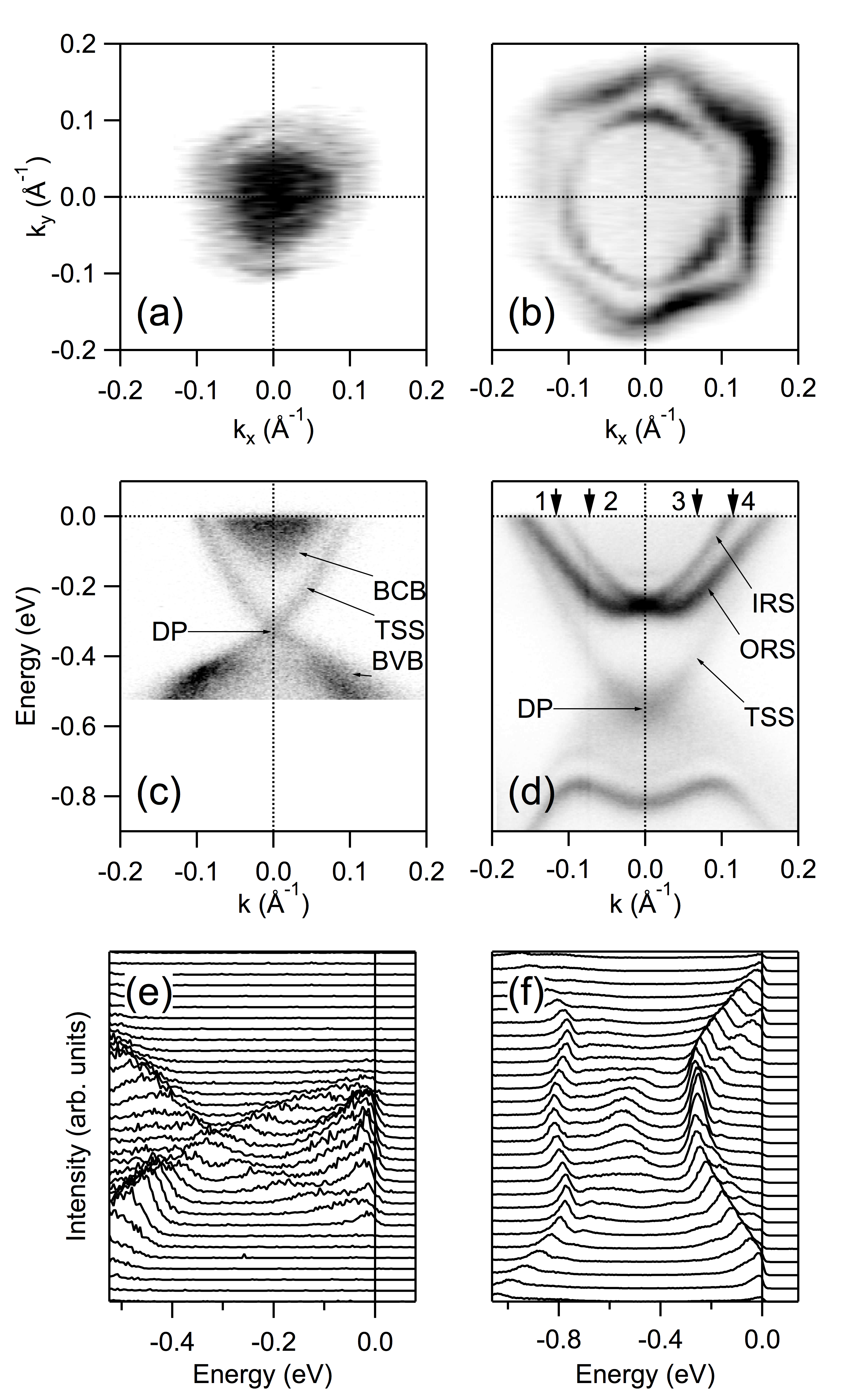}
\caption{Spin integrated ARPES of Bi$_2$Se$_3$. (a) Fermi Surface, (c) spectral intensity near the $\Gamma$M line in the Brillouin zone and (e) corresponding Energy distribution curves of the pristine sample. The topological surface state (TSS), the bulk conduction band (BCB), the bulk valence band (BVB) and the Dirac point (DP) are indicated in (c). (b), (d) and (f) show the corresponding spectra from the Rb-doped surface. Inner Rashba state (IRS) and outer Rashba state (ORS) are indicated in (d). The arrows (1-4) represent the approximate positions for the spin-resolved measurements. The spectra were measured using 18.5 eV photon energy at 20 K.}
\label{Fig1}
\end{center}
\end{figure}

Fig. \ref{Fig1}(c) shows the ARPES intenstiy from the pristine Bi$_2$Se$_3$ sample, measured along the momentum cut near the $\Gamma$M line in the surface Brillouin zone. The rapidly dispersing conical band represents the TSS with the Dirac point at -0.33 eV. Its FS is nearly a circle with the radius of $\approx 0.1$ \AA\ (Fig 1(a)). The bulk conduction band indicates electron doping due to Se vacancies, similar to previous studies.\cite{Xia2009} Adsorption of Rb on the surface of Bi$_2$Se$_3$ (Fig. \ref{Fig1}(b),(d) and (f)) induces additional electron doping  of the TSS, evident from the down-shift of the Dirac point and the corresponding enlargement of its FS that acquires a pronounced hexagonal warping.\cite{Valla2012a} However, this is not the only effect of doping: new states are also being formed inside the TSS cone and progressively filled with electrons donated by the adsorbed Rb. These new states always come in pairs, resembling the spin-orbit split Rashba states previously observed on Au(111) \cite{LaShell1996,Muntwiler2004} and Bi/Ag(111).\cite{Ast2007} The pair from Fig. 1(d) consists of two parabolic-like bands, displaced in momentum in a Rashba-like manner, intersecting at a new Dirac 
point at the zone center. At the highest doping levels, the outermost state is almost degenerate with the TSS, forming the FS nearly equal in shape and size (Fig. 1(b) and (d)). Its 
inner counterpart is significantly smaller, retaining the perfectly circular Fermi surface, even at the highest doping.
\begin{figure}[htb]
\begin{center}
\includegraphics[width=8cm]{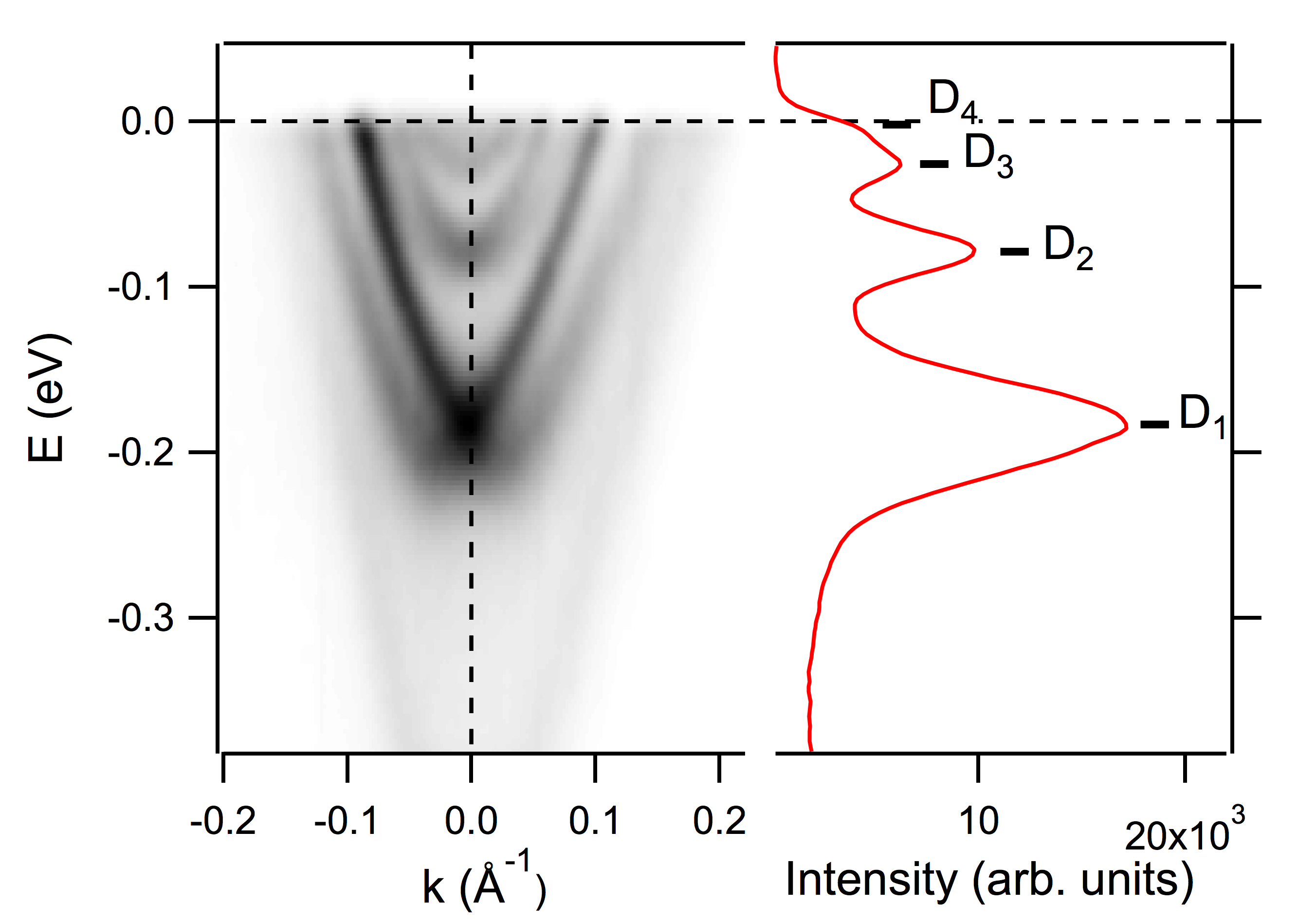}
\caption{Heavily Cu-doped surface of Bi$_2$Se$_3$. Spectral intensity along the $\Gamma$K line in the Brillouin zone (left) and the Energy distribution curve at the $\Gamma$ point, recorded at 35 eV photon energy. Multiple Rashba doublets are visible, with their degeneracy  points marked.
}
\label{Fig2}
\end{center}
\end{figure}

Similar phenomena have been also observed on Bi$_2$Se$_3$ covered with Cu (Fig. 2), Cs, K, Fe, Gd, water and CO, indicating that these states are intrinsic to the Bi$_2$Se$_3$ surface.\cite{Benia2011,Bianchi2011,Zhu2011,Valla2012a} The role of the adsorbate is to provide electron-doping, pushing these higher laying states below the Fermi level.
Indeed, slab calculations show that the paired states are quantum well states, confined in the potential well formed by the bulk gap and the surface potential and that they are empty when the surface is clean, but become partially occupied when electron donors are adsorbed on the surface.\cite{Yazyev2010,Zhu2011} The binding energy of these states is determined by the effective potential of the quantum well and by its \lq\lq{}thickness\rq\rq{}. 
Accordingly to the Rashba model, the spin degeneracy of these states is lifted by the spin-orbit coupling
\begin{equation} \label{eq:so}
\hat{H}_{\rm SO}=\frac{\hbar}{4 m_0^2c^2} \hat{\bf p} \cdot (\hat{\bm \sigma} \times \nabla V({\bf r})),
\end{equation}
in the presence of a breaking of the inversion symmetry at the surface. Here $m_0$ is the free electron mass, $\hat{\bm \sigma}=(\hat{\sigma}_x,\hat{\sigma}_y,\hat{\sigma}_z)$ is the vector of the Pauli matrices, and $V({\bf r})$ is the electric potential.
The major portion of the spin-orbit coupling occurs on the crystal side of the potential well, where the electrons feel the core potential of heavy Bi atoms. The observation that the Rashba splitting is larger for lower states therefore indicates that their wave functions extend deeper in the crystal, while the higher states have more weight on the vacuum side of the well. A similar behavior has been reported for the conventional image states.\cite{McLaughlan2004}

We note that only the spin-resolved studies can provide a definite proof that the observed states are Rashba states and determine their spin configuration relative to the TSS\rq{} one.
Therefore, we have performed the Spin-resolved measurements at four $k$ locations, indicated in Fig. \ref{Fig1}(d), at a relatively low photon energy (18.5 eV) to maximize both the energy and momentum resolution. In these conditions, the inner Rashba state has a somewhat weaker intensity than its outer counterpart (see Fig.\ref{Fig1}(f)).
\begin{figure}[tbp]
\begin{center}
\includegraphics[width=8cm]{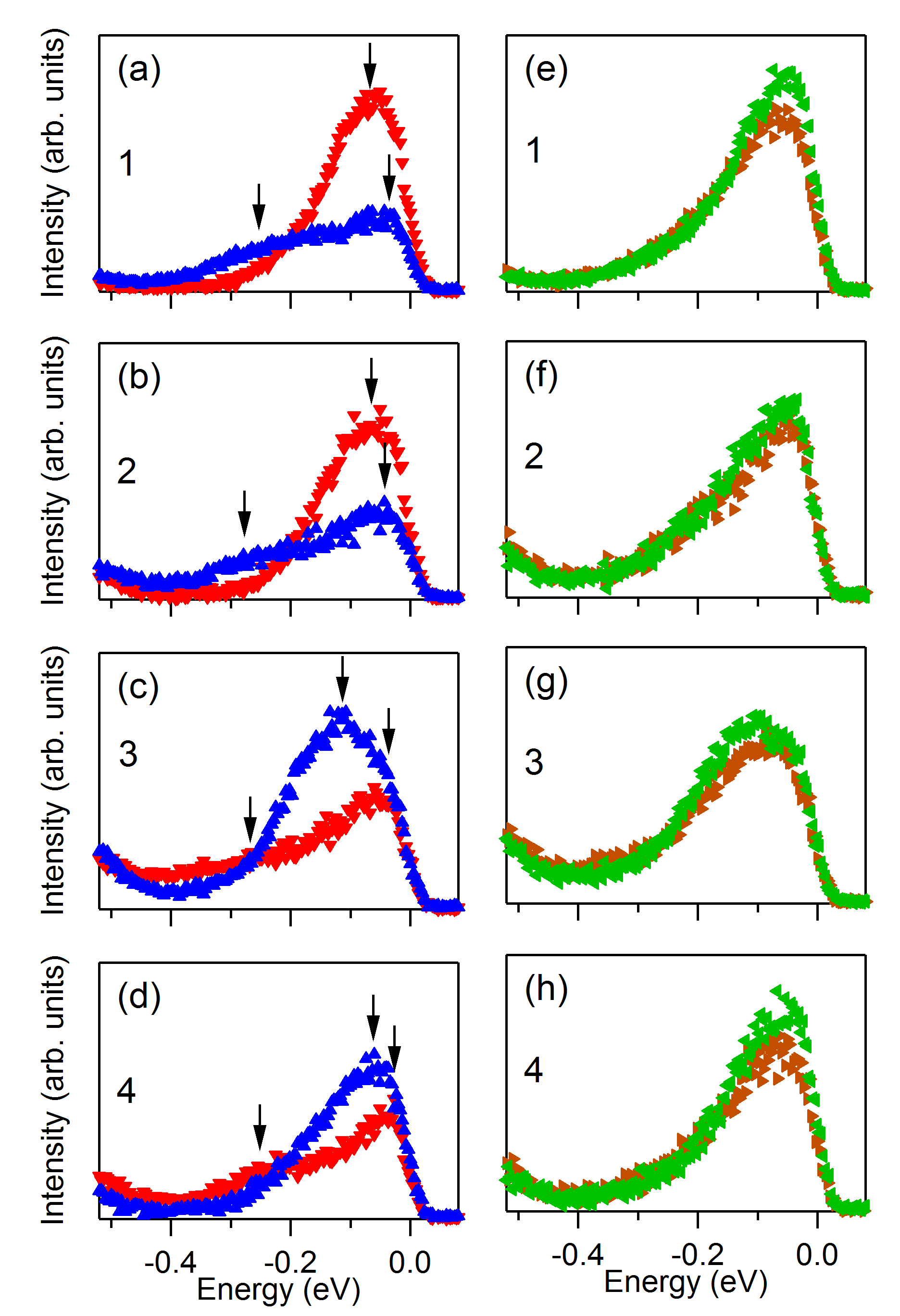}
\caption{Spin-resolved photoemission spectra of Bi$_2$Se$_3$ covered with Cu. (a)-(d) Spin resolved EDCs for spin component in $y$ direction. The arrows indicate peak positions. The corresponding $k$ positions (1-4) in the Brillouin zone are marked in Fig.\ref{Fig1}(d). (e)-(h) Spin resolved spectra for spin component in $x$ direction. The spectra were measured at 18.5 eV photon energy at T = 30K. 
}
\label{Fig3}
\end{center}
\end{figure}
The four detectors (Left, Right, Top and Bottom) of the Mott-polarimeter are oriented so that the asymmetry (A) in the Left-Right intensities reflects the spin polarization in $y$ direction, while the asymmetry in the Top-Bottom intensities reflects the polarization in $x$ direction. The spin polarization is obtained as P=A/S where S represents the Sherman function, or the efficiency of the spin polarimeter, S = 0.17 in our case. 

\begin{figure}[htb]
\begin{center}
\includegraphics[width=8.5cm]{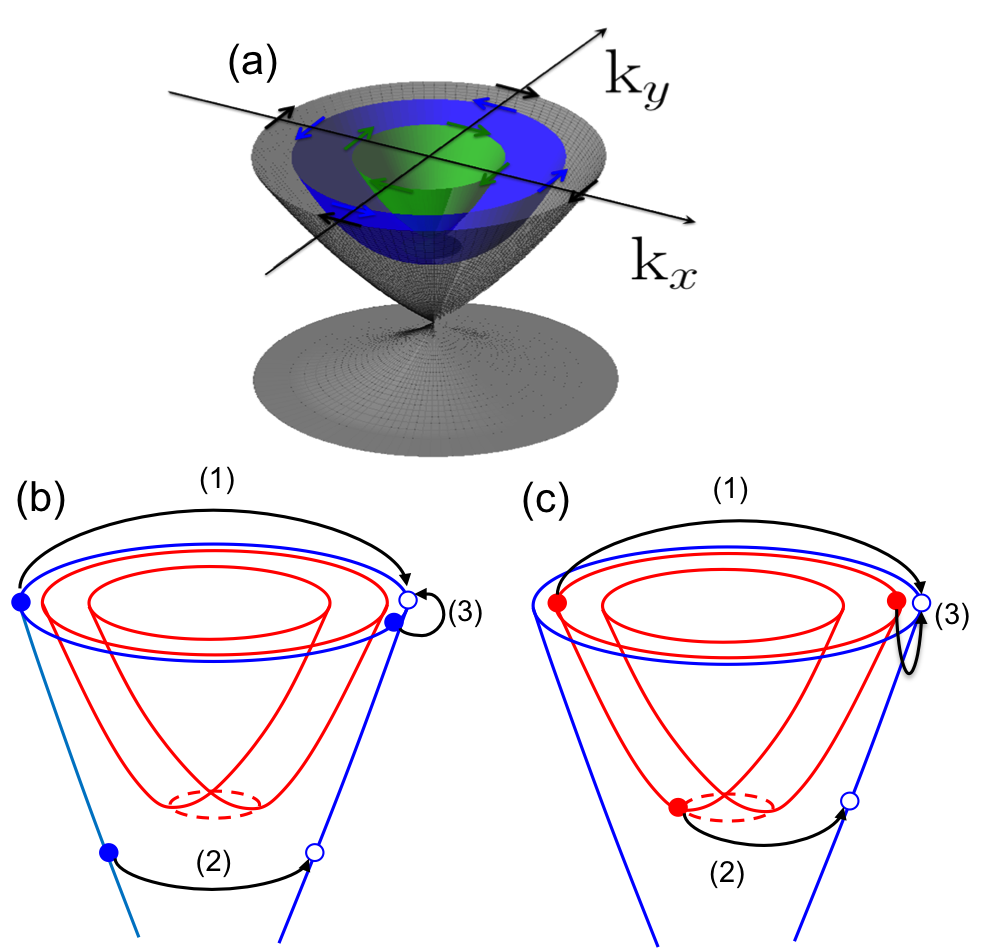}
\caption{Spin structure and elastic scattering on the doped Bi$_2$Se$_3$ surface. (a) In-plane spin structure of the TSS (gray), outer Rashba state (blue) and inner Rashba state (green). The arrows represent the in-plane spin direction of each band. (b) Schematic representation of  intra- and (c) inter-band scattering channels. (1) and (2) represent large-angle scattering events, while (3) is an example of small-angle event. (2) in (c) panel represents a threshold inter-band scattering event. Depending on the type of impurity, some of these events may be allowed or forbidden.
}
\label{Fig4}
\end{center}
\end{figure}

Panels (a)-(d) of Fig.\ref{Fig3} show the spin resolved spectra for the spin component in $y$ direction at the four $k$ points indicated in Fig. 1(d).
Blue and red spectra represent the EDCs for spin-up and spin-down, respectively.
At each $k$ point, there are three peaks marked by arrows: the lowest energy peak corresponds to the TSS, the next one is the outer Rashba state (ORS) and the peak closest to the Fermi level is the inner Rashba state (IRS). As expected, the sign of the spin polarization of the TSS inverts going from $k_x<0$ to $k_x>0$, consistent with the previous reports for the pristine Bi$_2$Se$_3$ surface, where the Fermi surface of the TSS shows left-hand spin helicity.\cite{Hsieh2009a,Pan2011c} 
From the spectra in Fig.\ref{Fig3} it is also apparent that the new states at lower binding energies do indeed form a Rashba doublet, displaying opposite spin polarization and  spin-momentum locking. Furthermore, the overall spin ordering is such that the spins of these three states -- TSS, ORS, IRS -- alternate in a way that produces an L-R-L order of spin helicities of the corresponding Fermi surfaces, as shown in Fig. 4(a).

The spin-resolved spectra for the $x$ component of the spin are almost degenerate, indicating a very small but finite spin polarization in this direction (Fig. 3(e)-(h)). There are two possible explanations for the residual polarization in the $x$ direction. First, the sample alignment might not be perfect - a small tilt would cause a finite $k_y$ and consequently, a small spin polarization in $x$ direction.
Second, the hexagonal warping of the states that occurs at high electron doping may cause an out-of-plane component of the spin polarization, reducing the in-plane one.\cite{Fu2009} Hexagonal warping would also spoil the in-plane spin-momentum locking $\bf{S}\perp\bf{k}$, which could result in a finite spin polarization in the $x$ direction.
Indeed, recent SARPES studies have reported a large out-of-plane spin component of the TSS on the pristine Bi$_2$Te$_3$ surface, which shows much larger warping than the TSS on Bi$_2$Se$_3$.\cite{Souma2011} As the electron doped Bi$_2$Se$_3$ surfaces studied here also show a strong warping, the breaking of $\bf{S}\perp\bf{k}$ locking and a resulting $x$ component of the spin in Fig. 3(e)-(h) is not surprising.

We note that the measured spin polarization of both topological and Rashba states is much smaller than the polarization measured on the clean surface of Bi$_2$Se$_3$.\cite{Pan2011c} This is probably due to the large number of states and their partial overlap in energy and momentum.  In this situation, the experimental resolution may not be sufficient to clearly resolve all the states and the measured polarization is inevitably reduced. As can be seen in Fig. \ref{Fig1}(d) and Fig. \ref{Fig2}, the TSS is almost degenerate with the ORS at the Fermi level. Similarly, the IRS of each Rashba doublet is nearly degenerate with the ORS of the next higher doublet. Assuming that the spins alternate in the same sequence as in the first three states, the measured polarization will be reduced, and its determination will require much better experimental resolution.

In Fig.\ref{Fig4}, we summarize the in-plane spin structure of the electron doped Bi$_2$Se$_3$ surface. It would be interesting to consider the effects of such spin structure on the scattering rates. In particular, it would be important for the understanding of scattering mechanisms to search for signatures of new scattering channels opened by the electron doping in the experimentally measured scattering rates. Fig. \ref{Fig4}(b) and (c) illustrate examples of possible scattering events.

A high resolution ARPES represents an ideal probe of quasiparticle scattering rates. The intensity of photoelectrons in an ARPES experiment $I(\textbf{k},\omega)$ is proportional to the single particle spectral function, $I(\textbf{k},\omega) =|M|^2A(\textbf{k},\omega)f(\omega)$, where $M$ represents the matrix element linking the initial and final states, $f(\omega)$ is the Fermi distribution and $A(\textbf{k},\omega)$ is the single particle spectral function:
\begin{equation}
\mathrm{A}(\textbf{k},\omega)\propto \frac{\mathrm{Im}\Sigma({\bf k},\omega)}{[\omega-\epsilon_{\bf k}-\mathrm{Re}\Sigma({\bf k},\omega)]^{2}+(\mathrm{Im}\Sigma({\bf k},\omega))^{2}}
\end{equation}
The complex self-energy $\Sigma(\textbf{k},\omega )$ contains 
the effects of the many body interactions and $\epsilon_{{\bf k}}$ is the non-interacting dispersion. 
The real part, Re$\Sigma(\textbf{k},\omega)$, gives a shift in energy and associated mass enhancement, while the imaginary part Im$\Sigma(\textbf{k},\omega)$ gives the scattering rate of a quasiparticle. 

\begin{figure}[htb]
\begin{center}
\includegraphics[width=8cm]{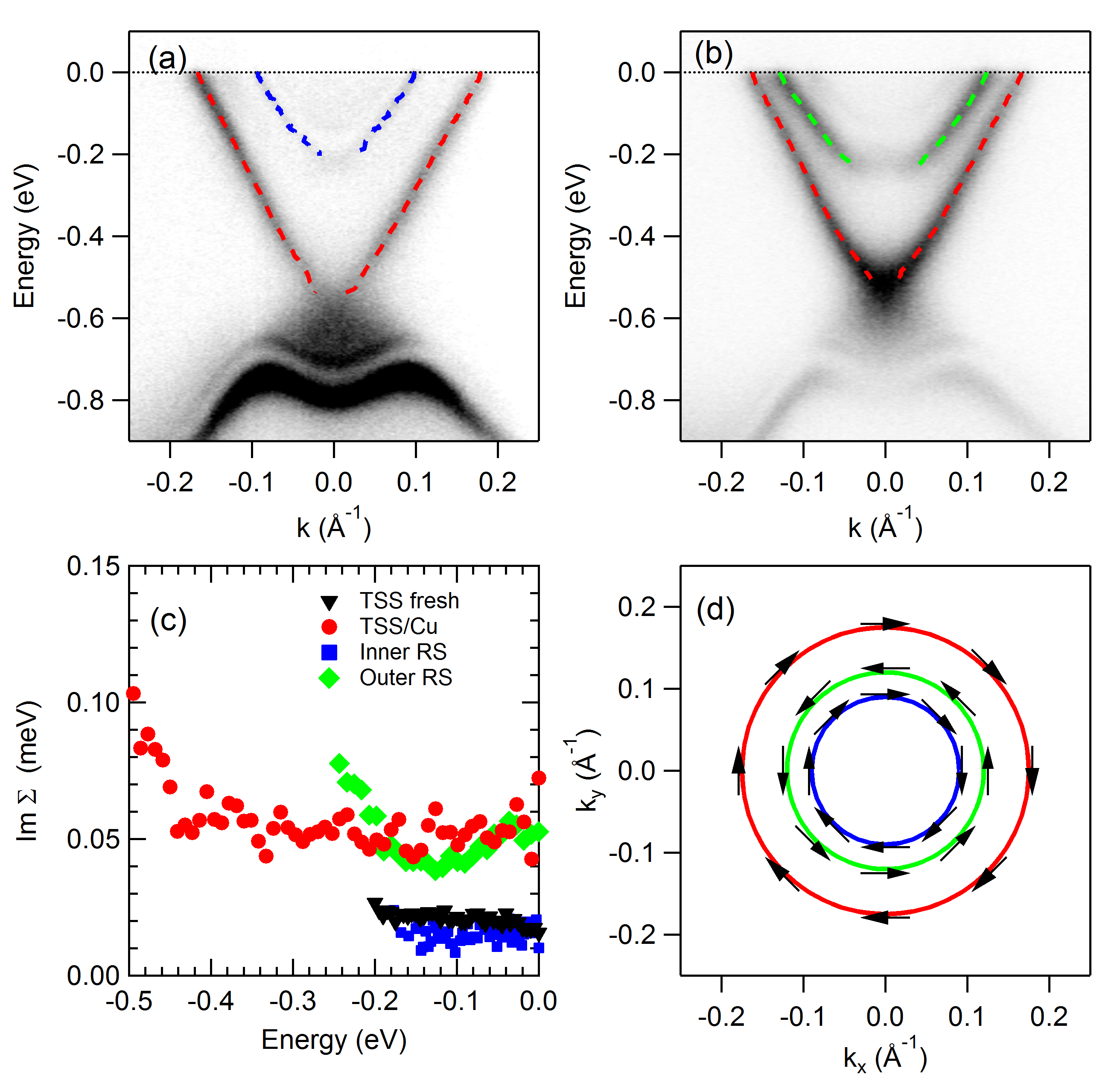}
\caption{The scattering rates of the topological surface state and the lowest Rashba doublet. (a) The ARPES spectrum of Bi$_2$Se$_3$ covered with Cu, measured with 60 eV photons. Red and blue dashed lines represent the dispersions of IRS and TSS, respectively. (b) The ARPES spectrum of the same surface, measured at 50 eV. The red and green dashed lines represent the dispersions of the ORS and the TSS, respectively. (c) The scattering rates of pristine (triangles) and heavily doped (circles) TSS, ORS (diamonds) and IRS (squares). All spectra were measured at T = 20 K. (d) Schematic representation of the surface spin structure.
}
\label{Fig5}
\end{center}
\end{figure}

As noted before, the TSS and ORS are nearly degenerate near the Fermi level (see Fig.\ref{Fig1}(b) and (d)). This, and the fact that at low binding energies there are multiple peaks in momentum distribution curves (MDCs),\cite{Valla1999a,Valla1999} makes the fitting of spectra recorded at 18.5 eV photon energy relatively unreliable. Fortunately, we have found that the photoemission cross-section of the TSS and Rasba doublets varies dramatically with photon energy and that at certain photon energies, some of the probed states could be almost completely suppressed, enabling much more reliable fitting of the remaining states. 
For example, with 60 eV photons, we see only the TSS and the IRS, as shown in Fig. \ref{Fig5}(a), while  at 50 eV, the spectra from the same sample show only the TSS and the ORS, as shown in Fig. \ref{Fig5}(b). Therefore, we have used these two photon energies to measure the self-energies of the TSS, IRS and the ORS. The scattering rates are determined from $\Gamma(\omega)=2|$Im$\Sigma(\omega)|=\Delta k(\omega)v_0(\omega)$, where $\Delta k(\omega)$ is the measured width of the Lorentzian-fitted peak in MDC and the bare group velocity $v_0(\omega)$ was approximated by the measured (renormalized) one, $v_m(\omega)$. The used approximation, $v_0(\omega)=v_m(\omega)$, implies Re$\Sigma\rightarrow 0$, consistent with the observation that Im$\Sigma$ is nearly $\omega$-independent for all three probed states. 

The results for Im$\Sigma(\omega)$ for the case of Bi$_2$Se$_3$ surface doped with Cu are shown in  Fig.\ref{Fig5}(c).
As can be seen, Im$\Sigma(\omega)$ of TSS for the heavily doped surface is larger than that of the freshly cleaved surface,\cite{Valla2012a} indicating that the TSS loses its coherence even in the presence of non-magnetic scatterers. 
For a given Cu coverage, the Im$\Sigma(\omega)$ of the ORS is very similar to the Im$\Sigma(\omega)$ of the TSS. Surprisingly, the Im$\Sigma(\omega)$ of the IRS is much smaller than the other two for the same Cu coverage and it is comparable to the Im$\Sigma(\omega)$ of the pristine TSS. 
We note that our previous studies did not find any significant difference in the scattering when the surface was covered with non-magnetic or magnetic adsorbates, for any concentration of adsorbates.\cite{Valla2012a} This suggests that the back-scattering ($q\sim2k_F$), which should be very different for the two types of adsorbates, is not the dominant scattering process, but that the measured widths are dominated by the small angle scattering. 

If indeed the small $\bf{q}$ scattering dominates, the question is whether the inter-band or intra-band processes are more important. If the inter-band scattering makes a major contribution, the outer Rashba state would be sharper than both the TSS and the IRS, as its scattering into the latter states would require a spin-flip event, as inferred from our measured spin-structure. However, the measurements show exactly the opposite: the IRS is sharper than the other two states. Furthermore, no anomaly in Im$\Sigma(\omega)$ of the TSS is seen at the threshold for inter-band processes at the bottom of the Rashba doublet at $\sim250$ meV below the Fermi level (see Fig. \ref{Fig5}(c)). This suggests that the small $q$ intra-band processes dominate the scattering rates of all three states. 

The remaining question is why the IRS is so much sharper than the other two states. We note that its Fermi surface is smaller and still circular, whereas the other two states form heavily warped Fermi surfaces.\cite{Valla2012a} This difference in shape (and size) is probably the main factor for the observed differences in the widths of the states. As already noted, warping can cause an {\it out-of-plane} spin component and the breaking of $\bf{S}\perp\bf{k}$ locking which relaxes the overall scattering restrictions. Consequently, the states with the more warped Fermi surfaces are expected to be broader. Also, it is expected that the Im$\Sigma$ of a warped state should be anisotropic. The heavily warped Fermi surface is convex and with zero out-of-plane spin component around the $\Gamma$M direction, while along the $\Gamma$K direction, it has a concave curvature and the largest out-of-plane spin component. Therefore, we anticipate that the Im$\Sigma$ would be minimal near the $\Gamma$M lines and maximal along the $\Gamma$K lines. The experiments that would measure the anisotropy of the scattering rates along the Fermi surface would be very important for a more complete understanding of scattering processes. 

In summary, we have demonstrated that the doublets of states that become partially filled when the surface of Bi$_2$Se$_3$ is doped with various electron donors are indeed spin-orbit split Rashba states. With the TSS, they form a staggered spin structure, with spin helicities of the resulting Fermi surfaces following the L-R-L... sequence, starting from the TSS. We have also found that the scattering rates on these Fermi surfaces are governed by the intra-band processes that depend mainly on the shape and size of the Fermi surface. 

The work at BNL was supported by the US Department of Energy, Office of Basic 
Energy Sciences, under contract DE-AC02-98CH10886. The work at MIT was 
supported by the US Department of Energy under Grant No. DE-FG02-07ER46134
ALS is operated by the US DOE under
Contract No. DE-AC03-76SF00098.
 Work at UConn was supported by ???

\bibliography{TI}   

\end{document}